\documentclass[9pt,twocolumn,twoside]{osajnl}

\journal{ao} 

\setboolean{shortarticle}{false}

\usepackage{stfloats}
\usepackage{graphicx}
\usepackage{caption}

\title{TiN-GST-TiN All-Optical Reflection Modulator for 2\textmu m Waveband Reaching 85\% Efficiency}

\author[1,2,\textdagger]{Md Asif Hossain Bhuiyan}
\author[1,\textdagger]{Shamima Akter Mitu}
\author[1,*]{Sajid Muhaimin Choudhury}

\affil[1]{Department of Electrical and Electronic Engineering, Bangladesh University of Engineering and Technology, Dhaka-1205, Bangladesh}

\affil[2]{Department of Computer Science and Engineering, BRAC University, Dhaka-1212, Bangladesh}

\affil[$\dag$]{The authors contributed equally to this work.}

\affil[*]{Corresponding author: sajid@eee.buet.ac.bd}

\begin{abstract}
 In this study, we present an all-optical reflection modulator for 2\textmu m communication band exploiting a nano-gear-array metasurface and a phase-change-material Ge\textsubscript{2}Sb\textsubscript{2}Te\textsubscript{5} (GST). The reflectance of the structure can be manipulated by altering the phase of GST by employing optical stimuli. The paper shows details on the optical and opto-thermal modeling techniques of GST. Numerical investigation reveals that the metastructure exhibits a conspicuous changeover from $\sim$ 99\% absorption to very poor interaction with the operating light depending on the switching states of the GST, ending up with 85\% modulation depth and only 0.58 dB insertion loss. Due to noticeable differences in optical responses, we can demonstrate a high extinction ratio of 28 dB and a commendable FOM of 49, so far the best modulation performance in this wavelength window. In addition, real-time tracking of the reflectance during phase transition manifests high-speed switching expending low energy per cycle, on the order of sub-nJ. Hence, given its overall performance, the device will be a paradigm for the optical modulators for upcoming 2\textmu m communication technology.
\end{abstract}

\setboolean{displaycopyright}{true}

\begin{document}

\maketitle

\section{Introduction}
The demand of internet is ever increasing since its invention and it is considerably higher since the time of covid crisis. Existing internet communication media is dealing with the ongoing bandwidth requirement by incorporating time, polarization, and wavelength division multiplexing or high-spectral efficiency coding \cite{Richardson2010} or spatial mode multiplexing \cite{8710274}. Compared to other communication media, optical fiber (OF) has been broadly used in internet communication as it provides low transmission loss, high security, and wider bandwidth\cite{gangwar2012optical}. The present industrial telecommunication bandwidth resides between 1.3 and 1.6 \textmu m range. However due to the establishment of the digital signal processing assisted coherent detection technique, the theoretical capacity of current single-mode fibers for internet communication is approaching its limit, perhaps causing a "capacity crunch" sooner than anticipated \cite{5210179}. 
For the upraising application of the emerging health technology such as telemedicine, artificial intelligence, Internet of Things, big data, 5G communication technology, and day-to-day activities including online transactions, online education and online shopping \cite{sudhir2020adapting,byrnes2021communication}, existing fiber-based telecommunication system will no longer be able to meet up the future demand. \cite{ellis2011capacity}. This problem can be reduced momentarily by compressing the transmission data or using multiple parallel link communication. But a promising solution is to shift the operating wavelength at 2 \textmu m window\cite{ackert2015high}. P. J. Roberts et. al. \cite{roberts2005ultimate} already designed a hollow-core photonic bandgap fiber (HC-PBGFs) with very low loss (0.13 dB/km at 1900 nm) in contrast to the traditional optical fiber which demonstrates 1 dB/km loss at 2 \textmu m wavelength. So the operating wavelength of telecommunication can be shifted from 1.5 \textmu m to 2 \textmu m which is more graceful than the existing solution. \cite{desurvire2011science}. 

Researchers have already started contributing to turning the scheme of the communication window relocation into a reality. In accordance with the modification of optical fiber, Eoin Russell et al. proposed Thulium Doped Fibre Amplifier (TDFA) \cite{russell2018development} for the preferred spectral region; its gain media can be excited even by the cost-efficient 1550 nm pump laser. Tm\textsuperscript{3+} and Ho\textsuperscript{3+} ions-based lasers have been reported to operate as the optical sources for 2 \textmu m communication band as well \cite{Scholle2010}. On top of that, GeSn/Ge multiple-quantum-well \cite{Xu:19}, GeSn on SOI substrate-based \cite{li202130}, single layer GeSn based resonant cavity enhanced \cite{chen2022transferable} photo-detectors have shown convincing performance at two-micron-wavelength. Additionally, to ensure a complete updated communication system for 2 \textmu m band, a similar focus should be imposed upon the optical modulator or switch which is one of the key components of the communication system, capable of transmitting data at the root level of data communication with a minimal error margin. Accordingly, Wei Cao et al. \cite{cao2018high} proposed silicon-on-insulator (SOI) based high-speed modulators that can operate at 1950 nm wavelength. Moreover, Weihong Shen et al. \cite{Shen:22} designed a silicon microring modulator for 2 \textmu m waveband. Nevertheless, silicon-based optical modulators operate either in resonant mode: limited to narrow band applications and susceptible to temperature variation \cite{Xu2005} or in non-resonant mode: requiring too long interaction length to be highly integrated \cite{Liu2004}. On the other hand, WbO\textsubscript{3} \cite{Franke2000}, graphene \cite{Zhong2020} and quantum well \cite{6112181} have also been studied for tunable and reconfigurable photonic devices leveraging external stimuli. Still and all, they are not power-efficient as they require a continuous power supply to sustain a state. One possible solution can be the incorporation of non-volatile phase-change-material (PCM) which can maintain its state reliably even during the absence of external stimuli and shows pronounced contrast in optical properties between its states consuming low switching energy.

\begin{figure*}[t]
    \centering
    \includegraphics[width=0.9\textwidth]{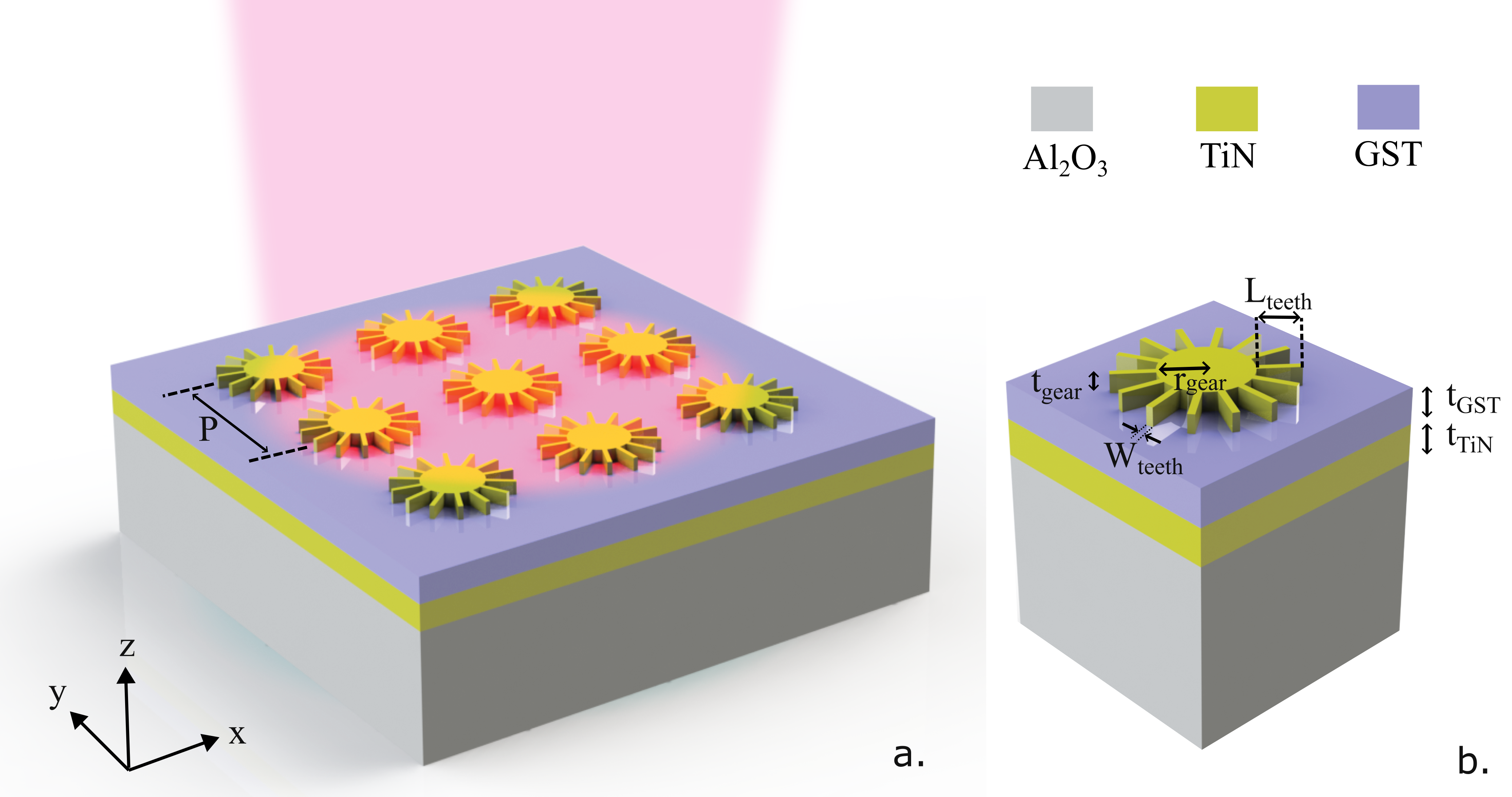}
    \caption{3D schematic illustration of (a) the proposed PCM based optical modulator and (b) one unit cell of the proposed structure demonstrating different structural parameters.}
    \label{fig1}
    \end{figure*} 

In recent years, due to unique combination of optical properties upon phase change \cite{zhang2018broadband}\cite{yu2017ultracompact}, many phase change materials such as Sb\textsubscript{2}S\textsubscript{3} \cite{Delaney2020}, Sb\textsubscript{2}Se\textsubscript{3} \cite{Hemmatyar2021EnhancedMU}, Ge\textsubscript{2}Sb\textsubscript{2}Se\textsubscript{4}Te\textsubscript{1} (GSST) \cite{Song:19}, Ge\textsubscript{2}Sb\textsubscript{2}Te\textsubscript{5} (GST) \cite{PrakashS:22} and VO\textsubscript{2} \cite{Badri_2022} have drawn more attention in the fields of optical absorption \cite{Pourmand2021}, modulation \cite{Kang2020}, filtration \cite{Badri:21} and micro-display \cite{Hemmatyar2021EnhancedMU}. Though VO\textsubscript{2} shows discernible metal-to-insulator phase transition, it is volatile. Conversely, germanium-antimony-tellurium (GST) is a non-volatile PCM material \cite{zhang2019nonvolatile} that has non-identical optical properties between its amorphous and crystalline phases and therefore, has the potentiality for optical modulation in nanophotonic metadevices \cite{cao2020photonic}. Also, its optical properties can be modified rapidly by applying electrical, thermal, or optical stimuli \cite{raeis2017metasurfaces}- a property that can be applied for designing a high-speed modulator. Moreover, GST demonstrates a decent endurance: about 10\textsuperscript{6} cycles (10\textsuperscript{11} in phase change memories \cite{Savransky2006}), which can be improved by doping (N, C, Zn) \cite{martin2022endurance}. However, for the desired spectral region, GST-based optical modulator is less-explored although rigorously studied for 1.5 \textmu m communication band \cite{Duan2022,Moradiani2021,Mahmoodi:21}.

Recent studies have indicated the possibility of a wide range of applications of all-optical modulator in the field of optical communication networks because of its staggering capability to transmit high-speed data over hundreds of kilometers without the requirement of any regeneration procedure \cite{minzioni2019roadmap}. All-optical modulator has become an essential component for the development of ultrafast, ultralow power consumption and flexible optical communication system \cite{li2018silicon,o2007optical}. \\

In this paper, we have designed a gear-shaped structure to develop an all-optical modulator for 2 \textmu m wavelength that exhibits a large extinction ratio, substantial modulation depth, low insertion loss, and a high figure of merit with the fast switching speed and sub-nJ switching energy. Firstly, we have described the structure of the proposed modulator followed by a theoretical model for its operation. Next, optical responses along with the switching mechanism of the device have also been discussed. Finally, we have compared the performance of our artifice to that of recently reported modulators.

\section{Structure}

The proposed GST-based modulator is made up of an opaque metallic TiN film atop a sapphire (Alumina) substrate and a TiN nanogear array above the TiN film separated by a layer of PCM demonstrated in fig. \ref{fig1}(a). Though many PCMs have been thoroughly studied and researched for a long, due to advantages of extreme scalability \cite{wang2018highly}, nonvolatile phase transition\cite{guo2019review}, high stability\cite{ding2019theoretical}, high switching speed\cite{lv2006electronic}, low optical losses in NIR region \cite{fang2021non}, quick response to stimuli in comparison with other PCMs and many other extraordinary properties, we choose GST alloy as the phase change material in our structure. The nanogear array is periodic along x and y directions and all nanogears share equal core diameter, teeth length, teeth number, and thickness. The teeth of a single nanogear are uniformly arranged in a radial pattern around the disk. The planar TiN film is used to prevent any transmission through the structure and hence, the optical response of the structure is not affected by the substrate. When the structure is illuminated from the top by a TM-polarized plane wave (propagation along the negative z-direction), incident light might either be reflected or absorbed by plasmon coupling depending on the phase of GST.

Au and Ag are the most commonly used materials for plasmonic applications due to their excellent plasmonic properties \cite{lee2006gold}. However, because of diffusive characteristics on the nanoscale,  Ag, and Au should be avoided in high-temperature applications such as GST phase transformation. TiN is a promising plasmonic material alternative to Au and Ag, especially for semiconductor compatible environments and high temperature applications. So, TiN is used as the metallic component in our suggested structure instead of Au or Ag. 

In order to obtain the optimized structure, optical response of the modulator has been investigated sweeping several structural parameters: disk radius (R\textsubscript{gear}), teeth length (L\textsubscript{teeth}), teeth width (W\textsubscript{teeth}),teeth number(N\textsubscript{teeth}), gear  thickness (t\textsubscript{gear}), period (P), GST thickness (t\textsubscript{GST}), and TiN film thickness (t\textsubscript{TiN}). These structural parameters have been tuned to achieve maximum Modulation depth, larger extinction ratio, and minimum insertion loss at a communication window of 2 \textmu m wavelength. Moreover, the distance between two consecutive teeth should be greater than 20 nm \cite{Zhang2004} to make the nanostructure practically fabricable. This limitation was taken under consideration while designing the proposed nanostructure. The numerical values of optimized geometrical parameters are charted in table \ref{tab:tab1}:

\begin{table}[h!]
\caption{Structural parameters of optimized modulator}
\centering
\label{tab:tab1}
 \begin{tabular}{||c c c c||} 
 \hline
 {\color{blue} Parameter} & { Value(nm)} & {\color{blue} Parameter} & {Value(nm)} \\ [0.5ex] 
 \hline\hline
 \textbf{r\textsubscript{gear}} & 130 & \textbf{L\textsubscript{teeth}} & 120 \\ 
 \textbf{t\textsubscript{gear}} & 50 & \textbf{N\textsubscript{teeth}} & 14 \\
 \textbf{t\textsubscript{GST}} & 100 & \textbf{W\textsubscript{teeth}} & 20 \\
 \textbf{P} & 700 & \textbf{t\textsubscript{TiN}} & 100 \\
  [1ex] 
 \hline
 \end{tabular}
\end{table}

\begin{figure}[t]
    \centering
    \includegraphics[clip=true,trim=0 0 0 0,]{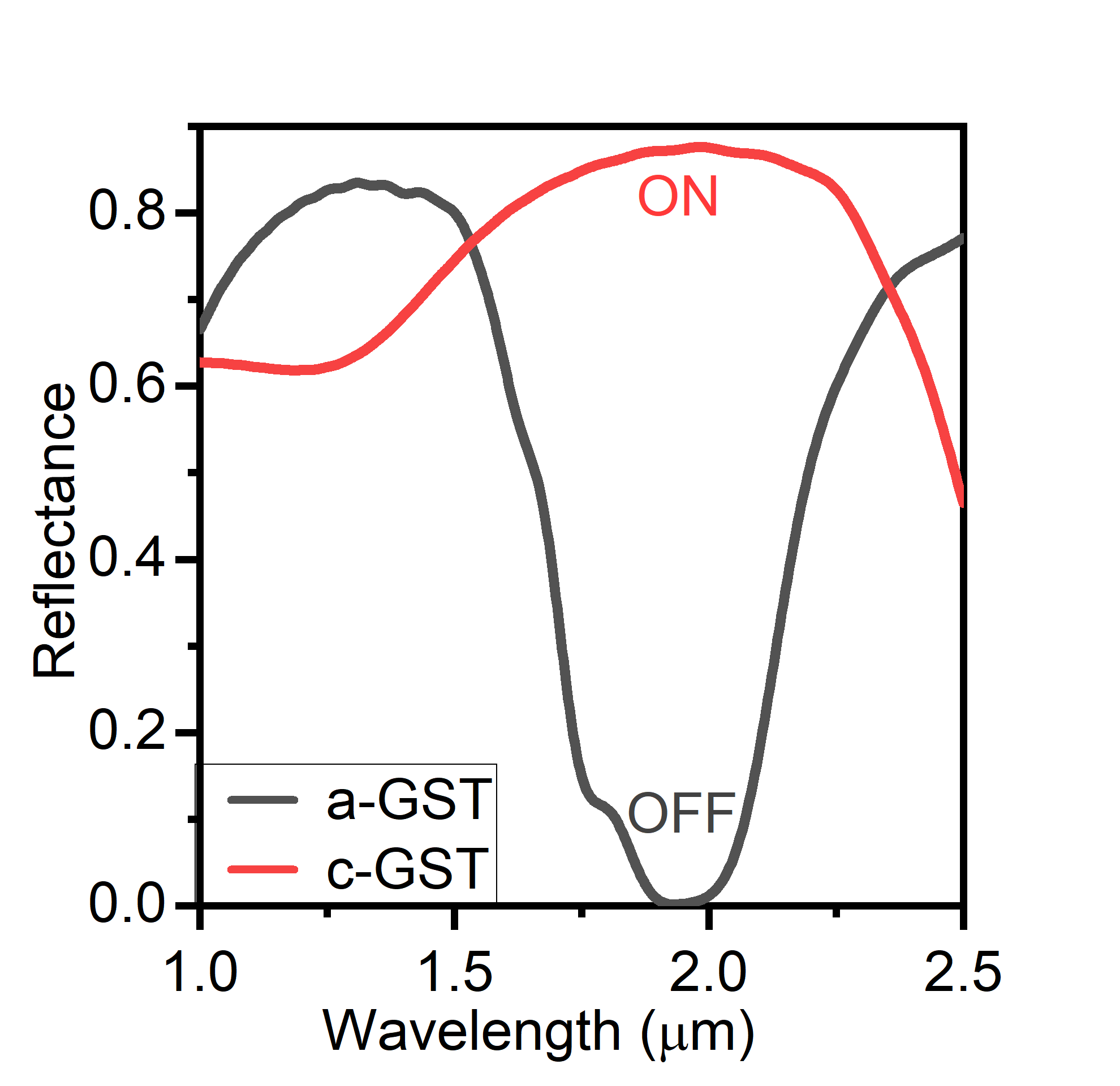}
    \caption{Reflection spectrum of the optimized structure. Most of the incident light of operating wavelength couple with the a-GST layer causing low reflection (OFF state). Conversely, the coupling is poor with the c-GST layer providing strong reflection (ON state).}
    \label{fig2}
\end{figure} 

\begin{figure}[t]
    \centering
    \includegraphics[width=0.45\textwidth]{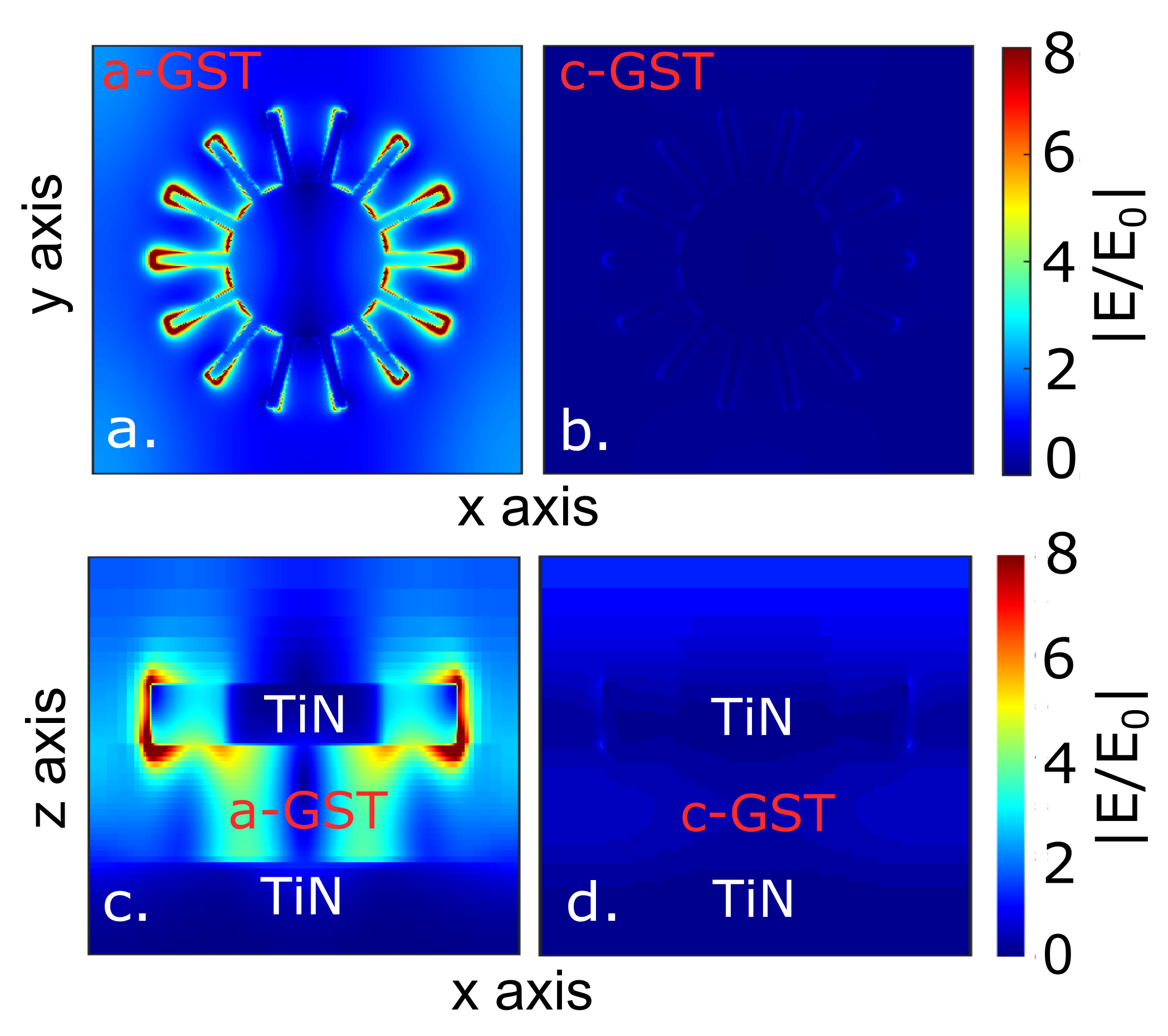}
    \caption{Electric Field distribution at the operating wavelength for the proposed nanostructure along the interface of GST and nanogear for (a) a-GST and (b) c-GST and along x-z cross-sectional plane going through the diameter of the gear for (c) a-GST and (d) c-GST. As depicted, strong resonance occurs when the GST is in amorphous state, but there is barely any field confinement inside the structure when a-GST is switched to c-GST.}
    \label{fig3}
\end{figure}

\section{Analysis Methodology}
We manipulate the quadruple resonance effect of previously reported nano-gear structure \cite{Sarker2020} by incorporating it with non-volatile phase change material – GST. As crystalline-GST (c-GST) has a larger refractive index than amorphous-GST (a-GST) in the operating wavelength window, the resonant wavelength demonstrates a red shift when the phase transition of GST occurs from amorphous state to crystalline state. The structural parameters are chosen in such a way that the nanogear couples incident light of 2 um wavelength with the underlying a-GST layer. This is referred to as the OFF state of the reflection based switching since the operational light is minimally reflected. To turn the switch ON, a laser pulse of sufficient power heats up a-GST above the crystalline temperature of GST to conduct the phase transition and as a result, the metastructure acts as a mirror for 2 um wavelength. The reflectance spectra have been investigated numerically by 3D FDTD method (See Supplement 1 for details) with periodic boundary conditions in x and y directions and perfectly matched layers (PMLs) in z directions of the unit cell depicted in fig. \ref{fig1}(b). The performance of the modulator has been evaluated by relevant performance benchmarks: modulation depth (MD), extinction ratio (ER), insertion loss (IL), and figure of merit (FOM) \cite{carrillo2016design}. The difference between maximum reflectance (R\textsubscript{cryst.}) and minimum reflectance (R\textsubscript{amph.}) of the modulator at operating wavelength is denoted as modulation depth.

\begin{equation}
    MD = R_{cryst.} - R_{amph.}
\end{equation}
Extinction ratio can be measured by the ratio of R\textsubscript{amph.} and R\textsubscript{cryst.} in dB \cite{carrillo2016design}.
\begin{equation}
    ER = -10log(\frac{R_{amph.}}{R_{cryst.}})
\end{equation}
Insertion loss has been calculated by the ratio of output optical power to input optical power in dB during ON state of the modulator. The obvious absorption of c-GST is accounted for by incorporating its extinction coefficient in the simulation. Finally, FOM is calculated by ratio of Extinction Ratio and Insertion loss during ON state\cite{das2021performance} .
\begin{equation}
    IL = -10log(\frac{P_{out}}{P_{in}})
\end{equation}
\begin{equation}
    FOM = \frac{ER}{IL_{on}}
\end{equation}
As a part of all-optical mechanism of the modulator, a super-Gaussian laser source \cite{Caiazzo2018} of 800 nm wavelength has been employed for thermal heating and subsequent phase transition of GST layer to conduct optical switching. The thermal power ($P_{th}$) absorbed by unit area of the structure in a single laser pulse is 

\begin{equation}
    P_{th} = I_{ov} \times F(r) \times exp(-\frac{(t-t_0)^2}{\tau^2})
\end{equation}
where $I_{ov}$ is the overlap integral of the laser spectrum (Power Spectral Density) and the absorbance profile of the modulator, $F(r)$ is the super-Gaussian portrayal of the incident beam in the spatial domain, $\tau$ is the time constant of the laser heating pulse and $t_0$ is the delay \cite{Cao2018}.

\begin{equation}
    F(r) = \frac{P_s}{\pi w^2}\times exp(-\frac{2r^n}{w^n})
\end{equation}

Here, $P_s$ is the maximum power of the laser source, $w$ is the radius of the beam, $r$ is the distance from the center of the beam and $n$ is the order of the Gaussian profile. Finally, the laser heating source $Q(r,t)$ has been designed to observe numerically the temperature distribution within the modulator structure i.e. the phase transition of the GST layer using the heat transfer module of a commercial software, COMSOL.

\begin{equation}
\label{laser}
    Q(r,t) = \frac{P_s}{\pi w^2}\times exp(-\frac{2r^n}{w^n})\times exp(-\frac{(t-t_0)^2}{\tau^2})
\end{equation}
The optical and thermal parameters of TiN and GST have been obtained from literature \cite{carrillo2016design,Cao2018,guo2019titanium,TAYLOR1964,Naik2013,zalden2014specific,Zhang2021,Xiong2011} (See Supplement 1).


\begin{figure}[hbt!]
    \centering
    \includegraphics[clip=true,trim=0 0 0 0,]{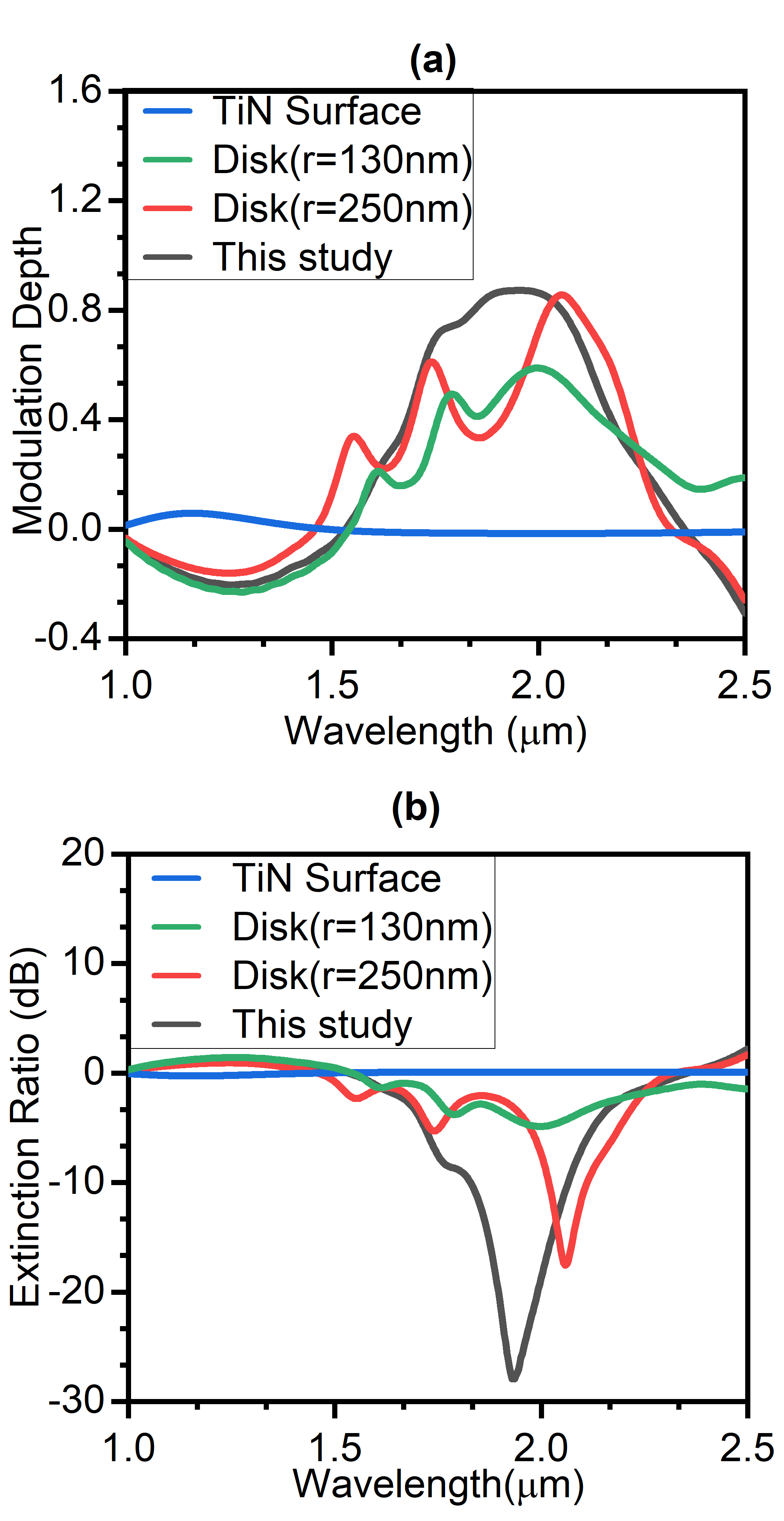}
    \caption{Comparison of (a) Modulation Depth \& (b) Extinction Ratio(dB) of different metasurfaces atop GST layer with our optimized gear-shaped grating.}
    \label{fig4}
\end{figure} 

\begin{table*}
\centering

\caption{Performance comparison of the proposed modulator with previously reported modulators}
\newcolumntype{x}[1]{>{\centering\arraybackslash\hspace{0pt}}p{#1}}
   
\begin{tabular}{m{2.5cm} x{1.7cm} x{1.55cm} x{1.3cm} x {0.6cm}x {1.7cm}x {1.3cm}x {1.5cm}x {1.8cm}}
 
\toprule
Device  & Switching \textcolor{black}{(exp./num.)\textsuperscript{\text{*}}} & Extinction Ratio(dB) & Insertion Loss(dB) & FOM & Modulation Depth&  Bandwidth & Bitrate (Gbps)& Operating Wavelength \\ \midrule


   { {\begin{tabular}[c]{@{}l@{}}Si\textsubscript{3}N\textsubscript{4}/VO\textsubscript{2} based\\ waveguide \cite{wong2019broadband} \end{tabular}}} & \begin{tabular}[c]{@{}c@{}}Optical \\ \textcolor{black}{(exp.)} \end{tabular}
   &
  6.72 &
  3.92 &
  1.71 & -
   &
   \textgreater{}100 nm&
  500 &
  1.5-1.6\\ \hline

  { {\begin{tabular}[c]{@{}l@{}}PCM embedded\\Si waveguide \cite{miller2017silicon} \end{tabular}}} & \begin{tabular}[c]{@{}c@{}} - \\ \textcolor{black}{(exp.)} \end{tabular}
   &
  14.9 &
  .6 &-
   &-
   &-
   &-
   &
  1.55 \\ \hline

{ {\begin{tabular}[c]{@{}l@{}}VO\textsubscript{2} controlled\\ Si waveguide \cite{markov2015optically} \end{tabular}}} & \begin{tabular}[c]{@{}c@{}} Electrical \\ \textcolor{black}{(exp.)} \end{tabular}
   &
  4.984 &
  7 &
  .71 &-
   &
  100nm &
   1&
  1.55 \\ \hline

{ {\begin{tabular}[c]{@{}l@{}} VO\textsubscript{2} integrated\\ plasmonic\\ waveguide \cite{moradiani2021high}\end{tabular}}} & \begin{tabular}[c]{@{}c@{}} Optical \\ \textcolor{black}{(num.)} \end{tabular}
   &
  18.7 &
  4.15 &
  - &
 -  &
   76&-
   &
  1.55 \\ \hline

{ {\begin{tabular}[c]{@{}l@{}}GST based\\ waveguide \cite{shadmani2019ultra} \end{tabular}}} & \begin{tabular}[c]{@{}c@{}} Electrical \\ \textcolor{black}{(num.)} \end{tabular}
   &
  \begin{tabular}[c]{@{}l@{}}8.12(TE)\\ 1.75(TM)\end{tabular} &
  \begin{tabular}[c]{@{}l@{}}1.15(TE)\\ 0.57(TM)\end{tabular} &-
   &-
   &
  500nm &
  0.4 &
  1.55\\ \hline
{ {\begin{tabular}[c]{@{}l@{}}VO\textsubscript{2} \& ITO\\ incorporated \\modulator  \cite{das2021performance} \end{tabular}}} &
  \begin{tabular}[c]{@{}c@{}} Electrical \\ \textcolor{black}{(num.)} \end{tabular} &
  2.925 &
  0.325 &
   8.98 &-
  &-
   &
   52.6&
  1.55 \\ \hline
{ {\begin{tabular}[c]{@{}l@{}}Plasmonic slot\\ waveguide using\\ VO\textsubscript{2} \& ITO \cite{amer2018vo2}\end{tabular}}} &
  \begin{tabular}[c]{@{}c@{}} Electrical \\ \textcolor{black}{(num.)} \end{tabular} &
  6.36 &
  1.13 &
    -&-
  &-
   &-
   &
  1.55 \\ \hline
  
{ {\begin{tabular}[c]{@{}l@{}}Hybrid \\ plasmonics VO\textsubscript{2} \\ modulator \cite{wong2017performance}\end{tabular}}} &
  \begin{tabular}[c]{@{}c@{}}Electrical/\\ Optical \\ \textcolor{black}{(num.)}\end{tabular} &
  7.6 &
  2.8 &
   2.7&-
   &\textgreater{}500nm & \begin{tabular}[c]{@{}c@{}} $\sim$1 (Electrical)\\ 
   \textgreater{}100 (Optical) \end{tabular}&
  1.55 \\ \hline
{ {\begin{tabular}[c]{@{}l@{}}VO\textsubscript{2} based \\ waveguide \cite{clark2017optically} \end{tabular}}} &
  \begin{tabular}[c]{@{}c@{}} Optical \\ \textcolor{black}{(num.)} \end{tabular} &
  14.76 &
  7.17 &
  - &-
   &-
   &-
   &
  1.55 \\ \hline

{ {\begin{tabular}[c]{@{}l@{}}Metallic grating\\ on ITO/GST \cite{carrillo2016design}\end{tabular}}} & \begin{tabular}[c]{@{}c@{}} Electrical \\ \textcolor{black}{(num.)} \end{tabular}
   &
  19.8 &-
   &-
   &
   .77&
   &-
   &
  1.55 \\ \hline

{ {\begin{tabular}[c]{@{}l@{}}\textbf{Proposed} \\ \textbf{Design} \end{tabular}}} & \begin{tabular}[c]{@{}c@{}} \textbf{Optical} \\ \textbf{\textcolor{black}{(num.)}} \end{tabular} 
   &
   \textbf{28} &
   \textbf{0.58} &
   \textbf{49}&
   \textbf{.85} &
   \textbf{\textgreater{}400nm} &
   \textbf{0.05} & \textbf{1.95} 
\\
\bottomrule
\multicolumn{3}{l}{*num. - numerical, exp. - experimental}


\end{tabular}
\label{tab:tab2}
\end{table*}

\section{Results and Device optimization}
\subsection{Optical Response of The Modulator: Switching}
The optical response of the structure has been explored for the oblique incident of operating light with an angle of 10 degree implying 'V' shaped optical input-output channels and the reflectance spectra of the modulator at the output channel for both the phases of GST have been demonstrated in fig. \ref{fig2}. As plotted, there is no reflected signal to travel via the output channel when the incident light is absorbed by a-GST layer. However, when the GST is in its crystalline phase, incoming light is mostly reflected and flows through the output channel. It lacks strong electric field confinement between nanogear and bottom TiN film which occurs while the phase of GST is amorphous as illustrated in fig. \ref{fig3}. During amorphous GST, the strong field confinement, shown in fig. \ref{fig3}(a) and \ref{fig3}(c), results in the anti-symmetric coupled state, as the electric fields at the top and bottom TiN layers are out of phases depicted by the polarization current distribution (fig. \ref{fig:FigS2} in supplement 1). Moreover, this dipolar resonance eventuates an oscillating magnetic dipole in each of the teeth, orthogonally oriented to the electric dipole. Furthermore, at the resonant wavelength, this quadruple resonance stimulates gap surface plasmon between teeth of two successive periodically arranged nanogears, resulting in zero reflectance \cite{Sarker2020}. Due to the metallic TiN plate at the bottom, zero transmission through the modulator can be achieved. Alaee et al.\cite{alaee2015magnetoelectric} presented that under these conditions with both electric and magnetic dipoles, near perfect absorption can be accomplished by correct scaling of the dimensions. On the contrary, during the crystalline phase of the GST, the coupling of 2 \textmu m wavelength light with the nanostructure is discernibly reduced as depicted in fig. \ref{fig3}(b) and \ref{fig3}(d) and this eventually weakens the magnetic dipole \cite{raeis2017metasurfaces}. For this reason, the incident light is reflected efficiently. Thus, the electric dipole coupled to the metallic film has been varied by altering the phase of the GST layer, i.e. in the amorphous GST, an oscillating magnetic dipole has been generated perpendicular to the electric dipole, whereas in the crystalline GST, the electric dipole and corresponding magnetic dipole have been significantly decreased.

We have optimized our design to achieve a modulation depth as large as $\sim85\%$, an insertion loss as low as $0.59$ dB, an extinction ratio of $-28$ dB, and a FOM of $\sim$ 49 that is so far the best modulation performance in the communication wavelength window of 2 \textmu m \cite{ackert2015high}. Although 
 the real fabricated transmission-based devices may experience apparent losses due to the transmission of the signal through the metallic nanostructure \cite{Melikyan2014}, omitting waveguide coupling loss, our reflection-based modulator shows minimal loss as the off-resonant reflected signal does not interact with the lossy metallic structure.
 We have also compared the performance of the modulator for normal incidence of the input signal with oblique incidence but there was no discernible difference. In the spectral region of 1900-2100 nm, this nanostructure yields acceptable MD, ER, IL and FOM. Additionally, we have compared the performance of the nanogear array with nanodisk arrays of the radius of 130 nm and 250 nm and also with a plane surface of TiN. Fig. \ref{fig4} (a) and (b) demonstrates how far the nanogear structure outperforms the nanodisk performance.

 \begin{figure}[t]
    \centering
    \includegraphics[clip=true,trim=0 0 0 0]{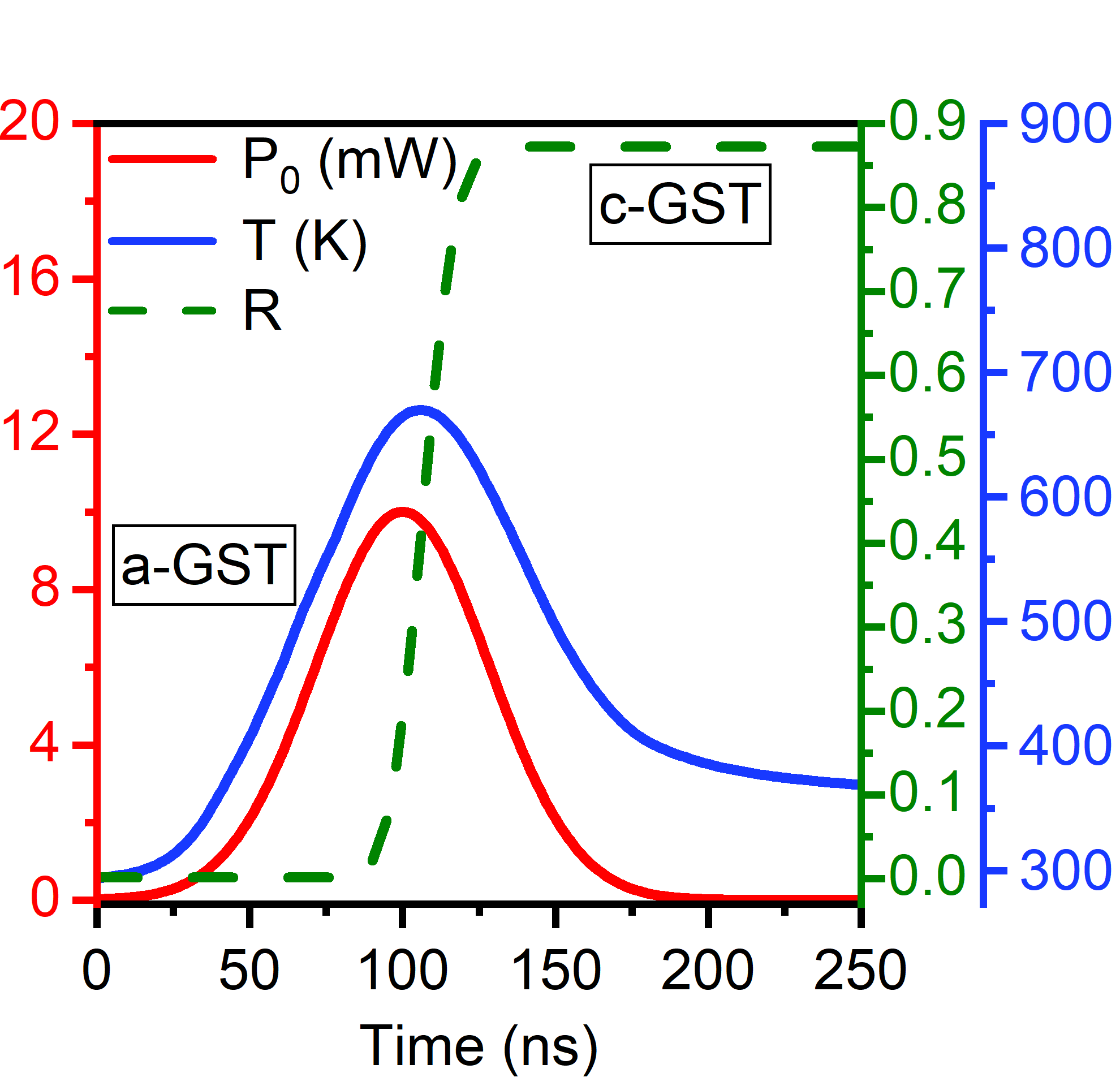}
    \caption{\label{f8} Heat Source power (P\textsubscript{0}), the temperature profile of GST layer (T), and real-time change in reflectance (R) manifest the crystallization of amorphous GST. When the temperature exceeds T\textsubscript{C} at around 85 ns, the reflectance of the structure begins to increase and continues elevating until the GST layer is fully crystallized at about 125 ns.}
\label{fig5}
\end{figure}

 \begin{figure}[t]
    \centering
    \includegraphics[clip=true,trim=0 0 0 0,]{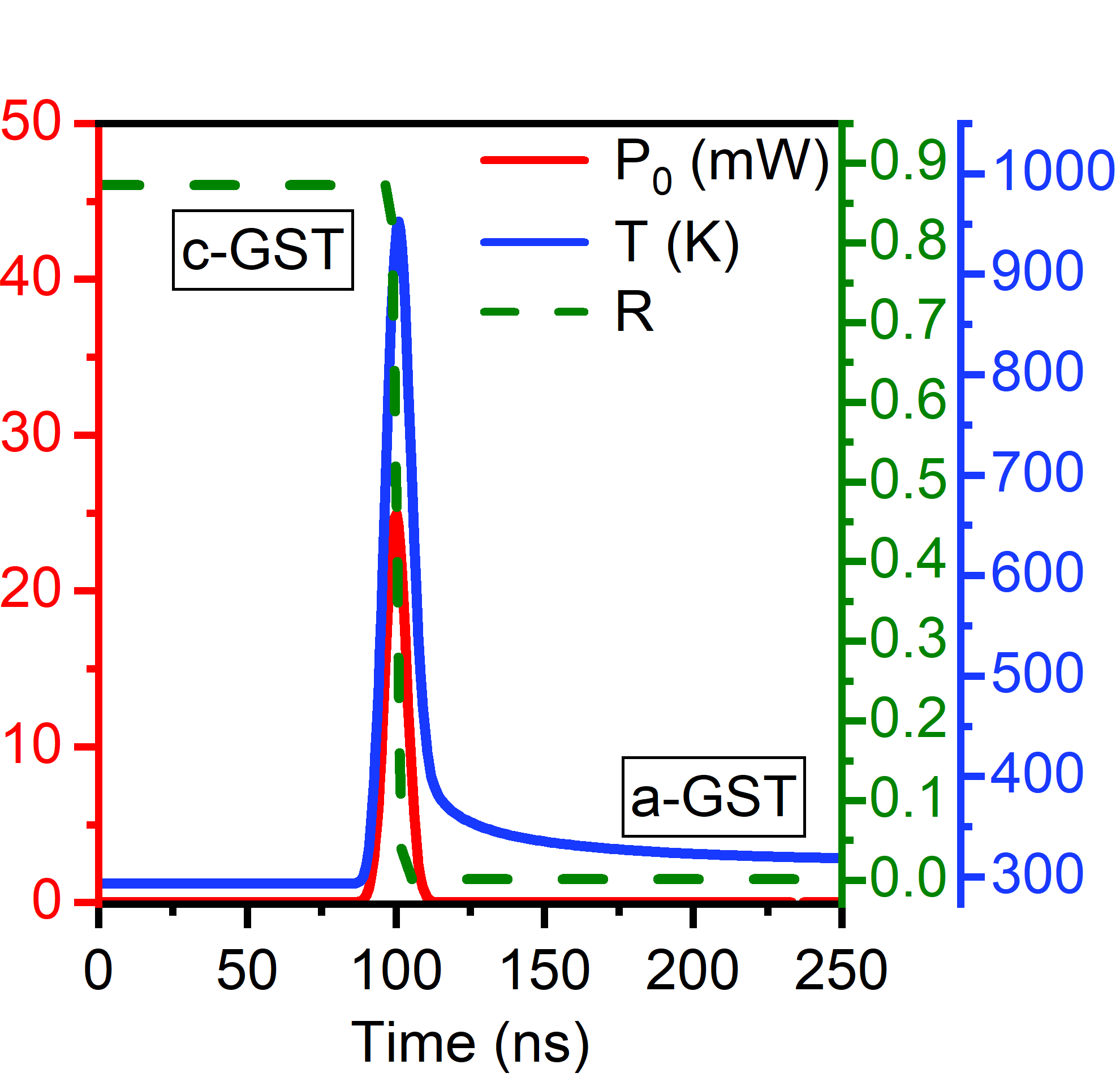}
    \caption{ Incident laser power (P\textsubscript{0}), the temperature profile of GST layer (T), and real-time change in reflectance (R) demonstrate the re-amorphization of GST. As soon as the temperature of the GST layer reaches T\textsubscript{M}, it melts and reflectance falls, as molten GST and a-GST have similar optical properties.}
    \label{fig6}
\end{figure} 

 \subsection{Opto-thermal Analysis of The Modulator: Phase Transition of GST} 
 The coupling between the operating light and the meta-structure has been modulated in this study by altering the phase of GST. Amorphous to crystalline phase transition can be accomplished by upraising the temperature of the GST layer above the crystallization temperature ($T_C=650K$, considering high heating rate \cite{Orava2012}) but not exceeding the melting point ($T_M=900K$ \cite{Akola2007}). The initial amorphous phase can be retrieved by melting the c-GST layer with rapid thermal heating and solidifying afterward. Both the heating processes can be facilitated by either electro-thermal heating or laser heating in PCM based devices \cite{Zhang2021_}. Two methods have been utilized in electro-thermal heating for a long: (i) electrical currents flowing directly through PCM and engendering Joule heating \cite{yu2017ultracompact} or (ii) resistive heating by external micro-heaters \cite{Abdollahramezani2022}. The former might lack uniform crystallization due to filamentation- a phenomenon of rising conductivity of PCM due to initial crystallization along the electric current pathway and allowing lower current through amorphous surroundings. Though the later strategy evades filamentation, micro-heaters require additional resistive materials \cite{Abdollahramezani2022,Taghinejad2021,Zhu2022} to generate heat. Herein, we incorporate optical heating through laser pulse, since uniformity can be achieved by increasing the area of illumination, and also, the optical absorption of GST around heating wavelength is not hindered by nanogear grating on top of the GST layer (See supplement 1 for the opto-thermal modeling of GST). Thus, we are capable of transforming the phase of the GST layer in this structure as fast as the bare GST layer by laser heating. For crystallization of the amorphous GST by elevating the ambient temperature of GST above T\textsubscript{C}, a super-Gaussian laser beam of 800 nm wavelength defined by eq. \ref{laser} with peak power, $P_s=$ 10 mW, beam radius, $w=$ 500 nm and time constant, $\tau=$ 40 ns has been projected on the modulator. Fig. \ref{fig5} demonstrates the input power of the beam, the simulated temperature profile of the GST layer as a function of time, and the concurrent variation of the reflectance. As we observe from the simulation, the temperature rises with input laser power, and the reflectance starts to go up when the temperature outstrips T\textsubscript{C}, implying that crystalline islands begin to develop at $\sim85$ ns. After $\sim125$ ns, the modulator is fully turned ON, indicating that the GST layer has completely crystallized. The energy required to turn ON the device is 1.445 nJ/\textmu m\textsuperscript{2}. By integrating P\textsubscript{0} (fig. \ref{fig5}) with respect to time and dividing the result by the area of simulated region, it is possible to determine the per-area energy associated with the crystallization process. For re-amorphization, the GST is melted momentarily by another laser pulse of $P_s=$ 25 mW and $\tau=$ 5 ns which ensures rapid thermal heating. Fig. \ref{fig6} illustrates the input laser power for phase retrieval and associated temperature distribution of the GST layer along with the descent of the reflectance: turning the modulator OFF. As depicted, the laser pulse causes hasty melting and subsequent quenching of GST allowing re-amorphization. The energy requirement is only 0.45 nJ/\textmu m\textsuperscript{2} in this process.

\begin{figure*}[t]
    \centering
    
    \includegraphics[width=0.9\textwidth] {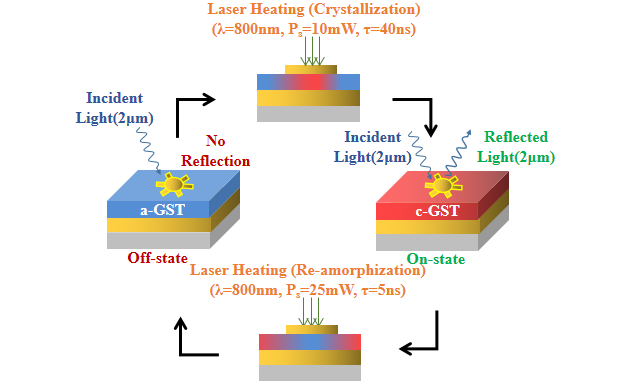}
    
    \caption{Schematic illustration of the working principle of our proposed GST based all-optical modulator for 2 \textmu m waveband.}
     \label{fig7}
    \end{figure*}

To capture the dynamic reflectance of this structure during GST phase transition, we have incorporated standard kinetics theory in our simulation \cite{doi:10.1021/acsphotonics.9b01456,Peng1997,Woods2017}. The amorphization is accomplished with a very short laser pulse and the modulator is turned OFF as soon as the GST layer melts, because molten GST has same optical properties as a-GST \cite{Peng1997,Nakayoshi_1992}. However, crystal formation in amorphous GST during turning the modulator ON is a bit slower process which acts as a speed-limiting factor \cite{Senkader2004}. Therefore, the performance of our phase-change device is mostly determined by the rate of crystallization of GST. As discussed earlier, from fig. \ref{fig5}, it can be inferred that a-GST takes $\sim$ 40 ns to be fully crystallized, and thus, switching speed of $25 MHz$ can be achieved. The comprehensive mechanism of our proposed modulator has been summarized in fig. \ref{fig7}.

Performances of recently reported electro-optical and all-optical modulators and switches with active PCM layer have been summarized in table \ref{tab:tab2}. Unspecified data are denoted by dash (-) in the table. In \cite{moradiani2021high}, Moradiani et al. theoretically and numerically reported an all-optical hybrid Au/VO\textsubscript{2}/Au plasmonic modulator. They used a probe-pump spectroscopy setup for changing the phase of VO\textsubscript{2}. They reported 18.7 dB extinction ratio and 4.15 dB insertion loss, with 0.9 \textmu m modulator length. HMK wong et al. \cite{wong2019broadband} experimentally demonstrated an ultra-compact (4 \textmu m modulator length) all-optical modulator, with 1.68 dB/\textmu m and 0.98 dB/\textmu m extinction ratio   and  insertion loss respectively. They used VO\textsubscript{2} as the active layer and achieved the FOM of 1.71. A GST-based dual polarization electro-optical modulator is proposed in \cite{shadmani2019ultra}. They reported that 1.15 dB \& 0.57 dB insertion loss and 0.12 dB \& 1.75 dB extinction ratio for TE and TM polarized light respectively. Two plasmonic electro-absorption modulators (EAMs) based on VO\textsubscript{2} and ITO are investigated in \cite{das2021performance}. They have achieved 0.651 dB/\textmu m IL, 5.85 dB/\textmu m ER, and FOM of 8.98 for VO\textsubscript{2} based EAM. The length of the device is 500 nm. So, for the overall device calculated IL and ER are 0.325 dB and 2.975 dB respectively. In \cite{wong2017performance}, HMK Wong investigated the performance of nanoscale VO\textsubscript{2} modulators using hybrid plasmonics. It yields ER = 3.8 dB/\textmu m and IL = 1.4 dB/\textmu m (FOM = 2.7) for 2 \textmu m long device. The device shows much better performance than other VO\textsubscript{2} based silicon wire waveguides. In \cite{clark2017optically}, they theoretically demonstrated all-optical Au/VO\textsubscript{2} nanostructure. It offers an extinction ratio of 28.56 dB/\textmu m and IL of  7.17 dB with a 550 nm device length. They used a pump light of 1050 nm wavelength for changing the state of VO\textsubscript{2}. In \cite{miller2017silicon}, they achieved a compact hybrid VO\textsubscript{2} based optical switch with 14 dB extinction ratio and 2 dB insertion loss through experimental calculations. In \cite{markov2015optically}, they designed an electro-optic modulator based on near field plasmonic coupling between Au nanodisk and VO\textsubscript{2} layer. It yields an 8.9 dB/\textmu m extinction ratio with 0.56 \textmu m device length. SGC Carrilo \cite{carrillo2016design} proposed a GST based electro-optical modulator with 77\%  modulation depth and 20 dB extinction ratio at 1.55 \textmu m operating wavelength. In \cite{cao2018high}, W Cao reported a high speed silicon modulators at 2 \textmu m operating wavelength. In \cite{amer2018vo2}, a hybrid plasmonic electro-optical modulator incorporating VO\textsubscript{2} and ITO is proposed. The device exhibits 6.36 dB ER and 1.13 dB IL at 1.55 \textmu m. In contrast, our proposed design shows a very high MD of 85\%, ER as high as 28 dB, only 0.58 dB IL \& FOM as large as 49 with 25 MHz switching frequency implying that our proposed modulator outruns other PCM devices in terms of overall performance.

\section{Conclusion}
We exploit the significant disparity of the optical characteristics between the amorphous and the crystalline GST at 2 \textmu m wavelength to develop an all-optical reflection modulator for future communication technology. A nanogear array, which was previously reported to be utilized for refractive index sensing at the same operating wavelength region, has been optimized for our modulator to achieve a pronounced resonance causing barely any reflection during the OFF state of the switch escalating the modulation depth to 85\%. The Switching mechanism of our proposed modulator has been investigated by phase change dynamics of GST in time scale applying laser heating pulses of sufficient power revealing a 25 MHz switching frequency. A high ON/OFF ratio of $\sim 10^{2.8}$, low insertion loss, and sub-nJ switching energy make it a reliable candidate for 2 \textmu m communication band.

\begin{backmatter}
\bmsection{Funding} Research Grant from Center for Advanced Research and Study (CASR), Bangladesh University of Engineering and Technology (CASR Meeting No. 339 , Resolution No. 62, Date:07/04/2021)

\bmsection{Acknowledgments} The author SAM wishes to acknowledge support from the Ministry of Science and Technology (https://most.gov.bd/), Bangladesh through the NST fellowship 2021-2022/MS (Physical Science Group) and from Bangladesh University of Engineering and Technology through the postgraduate fellowship.

\bmsection{Disclosures} The authors declare no conflicts of interest.

\bmsection{Data Availability} Data underlying the results presented in this paper are not publicly available at this time but may be obtained from the authors upon reasonable request.

\bmsection{Copyright Notices} 
The Author(s) agree that all copies of the Work made under any of the above rights shall prominently include the following copyright notice: “© 2022 Optica Publishing Group. One print or electronic copy may be made for personal use only. Systematic reproduction and distribution, duplication of any material in this paper for a fee or for commercial purposes, or modifications of the content of this paper are prohibited. 

The final version of this article is published in  \href{https://doi.org/10.1364/AO.470247}{Applied Optics Vol 61(31), pp9262-9270 (2022) \hspace{21pt} DOI:10.1364/AO.470247} 

The accepted version of this article is submitted in ArXiv for archival purpose.  

\bmsection{Supplemental document}
See Supplement 1 for supporting content. 
\end{backmatter}

\section*{Supplement 1: SIMULATION SETUP}

\subsection{Optical Response of The Modulator: FDTD Analysis}
\begin{figure}[b]
\centering
\includegraphics[width=1\linewidth, height=5 cm]{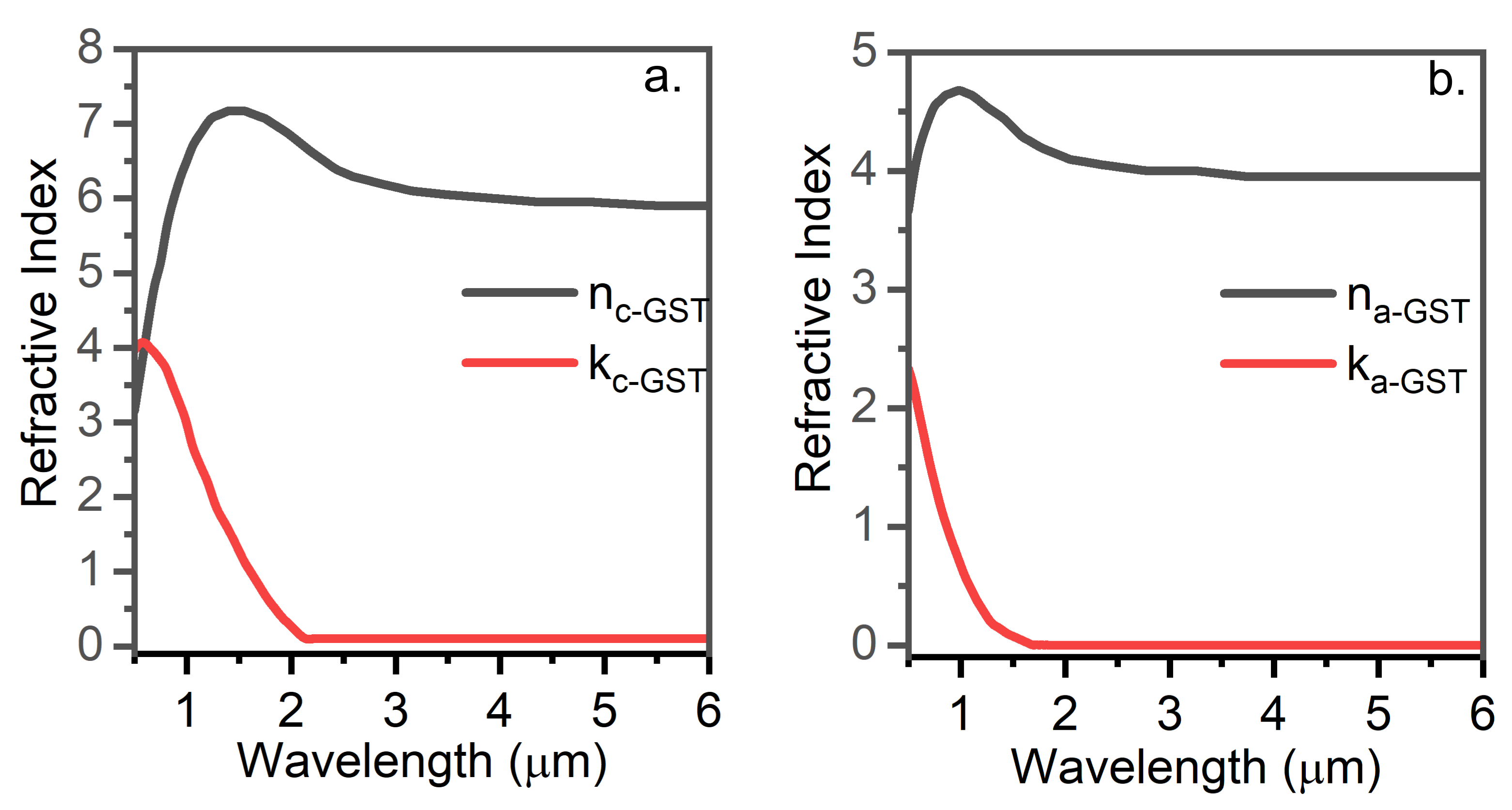}
\caption{Refractive index (n,k) data of (a) a-GST and (b) c-GST \cite{Cao2018}.}
\label{fig:FigS1}
\end{figure}

The optical response of the proposed nanostructure is analyzed numerically via the finite-difference time-domain (FDTD) method using the commercial software Lumerical FDTD Solutions. In the proposed modulator, GST layer is sandwitched between a TiN layer and a nano-gear array. We exploit the pronounced changeover of the optical properties of GST (as illustrated in fig. \ref{fig:FigS1}) upon phase change. The optical properties of TiN have been collected from \cite{guo2019titanium}. As our proposed structure can be tuned over a broad range of wavelength by changing the geometrical parameters like length of teeth (L\textsubscript{teeth}), number of teeth (N\textsubscript{teeth}), radius of disk (R\textsubscript{gear}), thickness of nanogear (t\textsubscript{gear}), thickness of metal film (t\textsubscript{TiN}) and thickness of GST layer (t\textsubscript{GST}), we have optimized our modulator for 2 um wavelength with a better performance (large modulation depth, high extinction ratio, low insertion loss and high FOM). The proposed nano-gear structure is illuminated by a TM polarized plane wave with 10 degree angle of incidence along the negative z direction. All numerical simulations (reflection spectra) are performed for a single cell of the nanostructure employing three dimensional (3D) FDTD method with periodic boundary condition in x and y directions and perfectly matched layer (PML) absorption condition in the positive and negative z direction. We have used different mesh size for different layers to increase the accuracy of simulation. For the disk, we have used 20 nm mesh size in all directions. Again, for teeth of the gear the mesh size has been set to 10 nm in x direction and 3 nm in both y and z directions. We have set a 'frequency-domain field and power' 2D monitor above the source and the source's wavelength is varied from 1 \textmu m to 2.5 \textmu m to obtain the reflectance spectra. To visualize electric field profile in a plane, we have placed the 2D monitor along the desired plane and simulated the structure for only the resonant wavelength of the source.

\begin{figure}[t]
\centering
\includegraphics[width=1\linewidth, height=5.5cm]{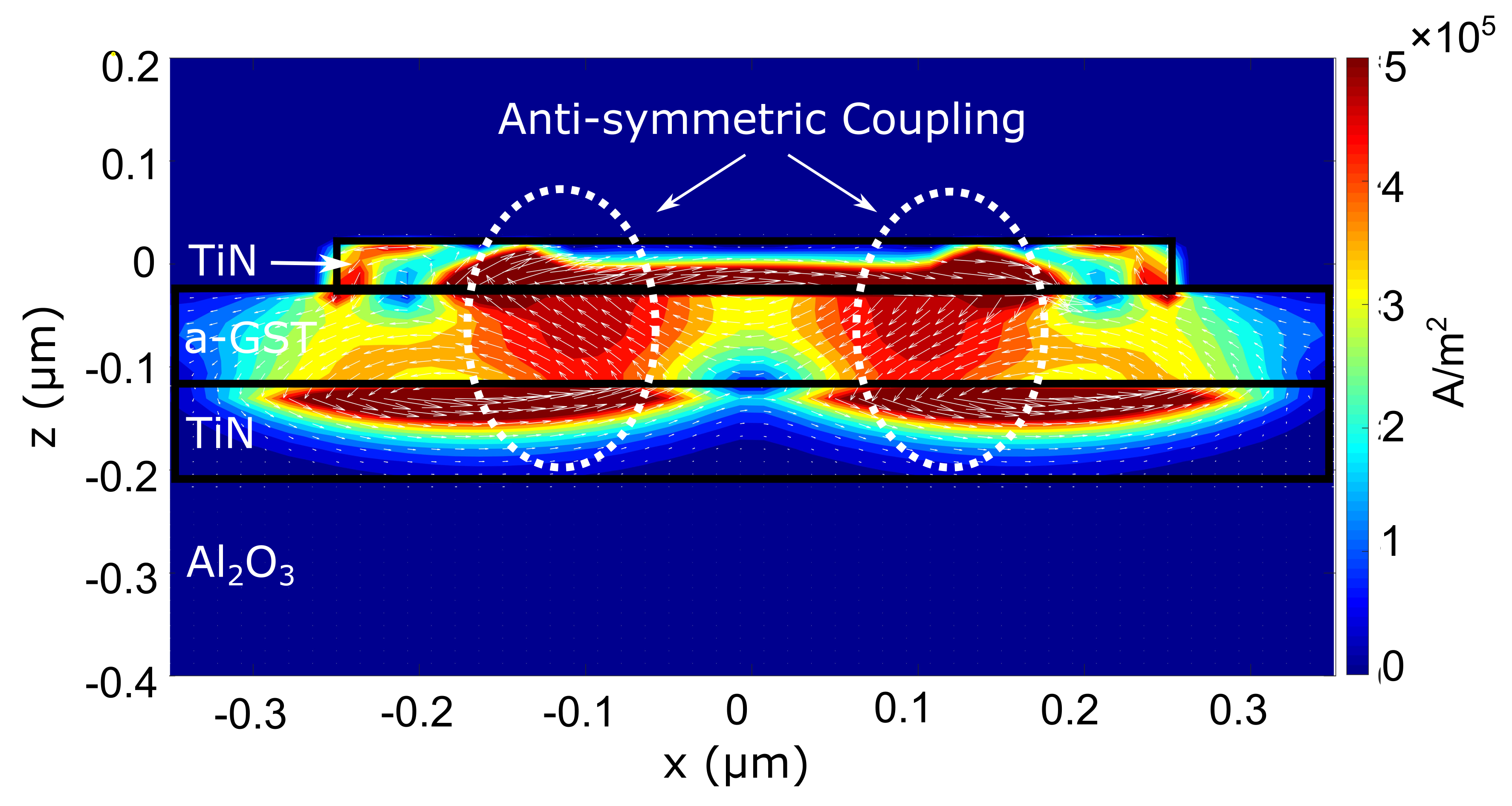}
\caption{Polarization current distribution during the OFF state (low reflectance) of the modulator along zx plane going through the diameter of the disk in nanogear. It is evident from the highlighted areas that polarization current either leaves GST layer at the top TiN layer and enters from the bottom or vice versa. This out-of-phase current distribution readily manifests the anti-symmetric coupled resonance. \cite{tang2011optical}}
\label{fig:FigS2}
\end{figure}

The TiN-GST-TiN structure produces anti-symmetric plasmon coupling between the top and the bottom TiN layers as depicted by the polarization current distribution in fig. \ref{fig:FigS2}. The polarization current direction is toward the GST layer in the bottom TiN layer, whereas the direction is outward the GST layer in the nanogear array, or vice versa.


\subsection{Opto-thermal Study: FEM Analysis}
The laser heating of the modulator and corresponding change in phases (ON/OFF) with temperature have been analyzed by the heat transfer module of COMSOL Multiphysics (FEM solver). In time-dependent 3D analysis, the module takes a heat source (rate of incident power per area), initial temperature, boundary conditions, and thermal properties of any structure as inputs and generates the temperature profile of the structure as a function of time. To do so, the module requires three physical properties of any material: density, specific heat, and thermal conductivity. Among the three materials (TiN, GST, Al\textsubscript{2}O\textsubscript{3}) forming the nanostructure, we used the built-in data for Al\textsubscript{2}O\textsubscript{3} and the thermal properties of other materials have been in Table \ref{tab:TiN} and Table \ref{tab:GST}. Some properties hardly vary with temperature while other properties possess a wide range of values for temperature variation. 

\begin{table}[htbp]
\centering
\caption{Thermal properties of TiN.}
\begin{tabular}{|l|l|}
\hline
\textbf{Property}             & \textbf{Values}                       \\ \hline
Density              & 5340 kg/m\textsuperscript{3} \cite{haynes2016crc}\\ \hline
Specific heat        &  plotted in fig. \ref{FigS3}             \\ \hline
Thermal conductivity &     27.3 W/(m.K) \cite{TAYLOR1964}         \\ \hline
\end{tabular}
\label{tab:TiN}
\end{table}


\begin{table}[htbp]
\centering
\caption{Thermal properties of GST.}
\begin{tabular}{|l|l|}
\hline
\textbf{Property}             & \textbf{Values}        \\ \hline
Density              & 6150 kg/m\textsuperscript{3} \cite{Zhang2021}\\ \hline
Specific heat        & plotted in fig. \ref{FigS4}             \\ \hline
Thermal conductivity & plotted in fig. \ref{FigS5}              \\ \hline
\end{tabular}
\label{tab:GST}
\end{table}

We have simulated a single cell of our reported structure to study its opto-thermal response. Physics-defined finer mesh has been selected for the simulation. The initial temperature has been set to 298 K and all the bare surfaces of the cell have been set to radiative-to-ambient mode. We have described the heat flux (laser heating beam) as eq. \ref{S1} and projected it from the top of the meta-surface. The heat flux is Gaussian in both the time and space domain. The power of the beam gradually decreases radially from the center of the beam. Also, the power gradually increases with time, reaches its peak at t=t\textsubscript{0}, and gradually falls again. The time-dependent study has been added, and the simulation time has been set to 300 ns with a time step of 0.1 ns. We have set a probe to the GST layer for the average temperature of the GST layer. 

\begin{equation}
\label{S1}
    Q(r,t) = \frac{P_s}{\pi w^2}\times exp(-\frac{2r^n}{w^n})\times exp(-\frac{(t-t_0)^2}{\tau^2})
\end{equation}

Here, $P_s$ is the maximum power of the laser source, $w$ is the radius of the beam, $r$ is the distance from the center of the beam, $n$ is the order of the Gaussian profile, $\tau$ is the time constant of the laser heating pulse, and $t_0$ is the delay. For crystallization, we set $P_s=$ 10 mW and $\tau=$ 40 ns for the laser beam which takes the temperature of the GST layer above T\textsubscript{C}. For re-amorphization, we put $P_s=$ 25 mW and $\tau=$ 5 ns to melt the GST layer momentarilly.

\begin{figure}[htbp]
\centering
\includegraphics[width=1\linewidth, height=7cm]{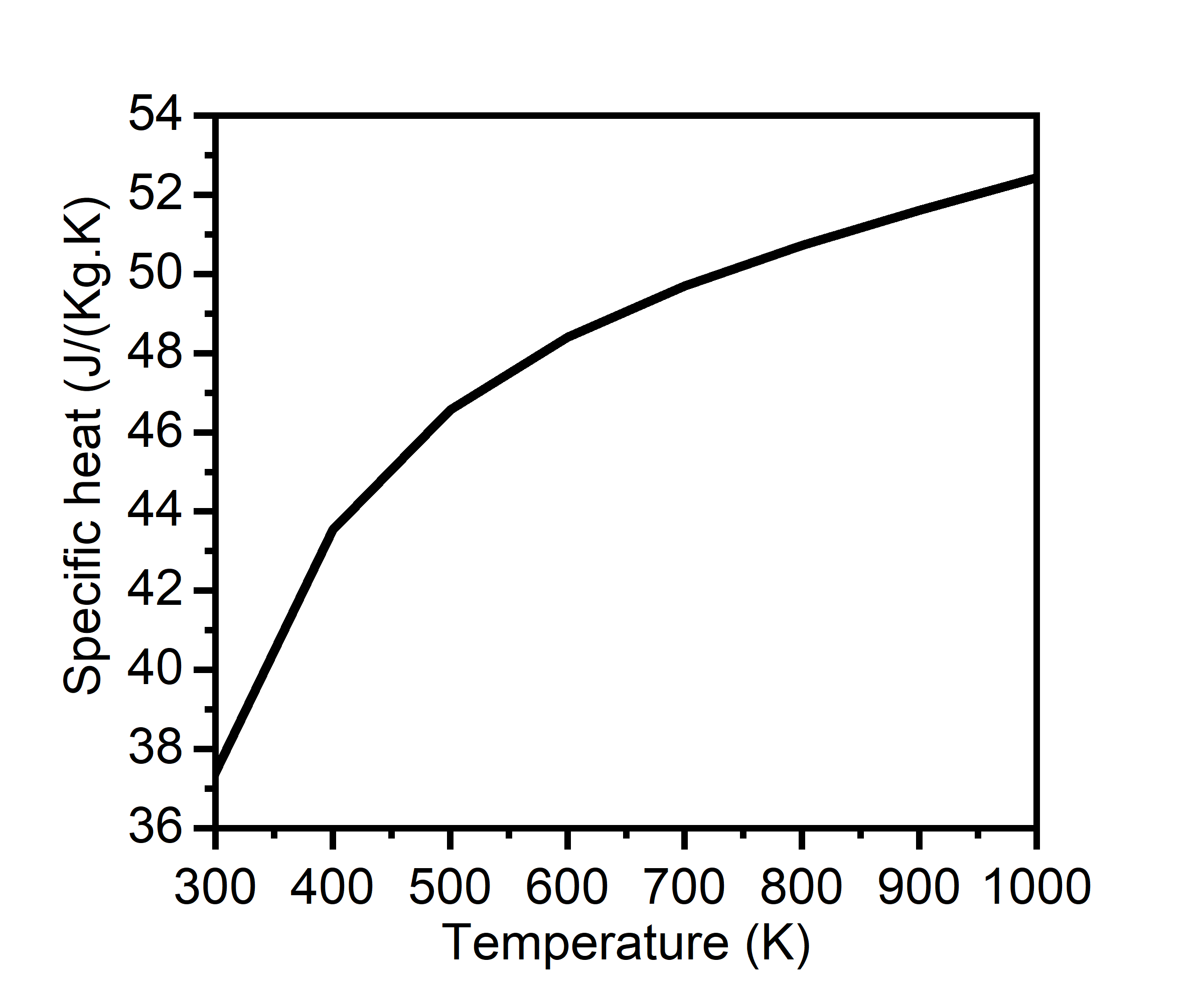}
\caption{Variation of specific heat of TiN with temperature provided by \cite{chase1998nist}.}
\label{FigS3}
\end{figure}

\begin{figure}[htbp]
\centering
\includegraphics[width=1\linewidth,height=7cm]{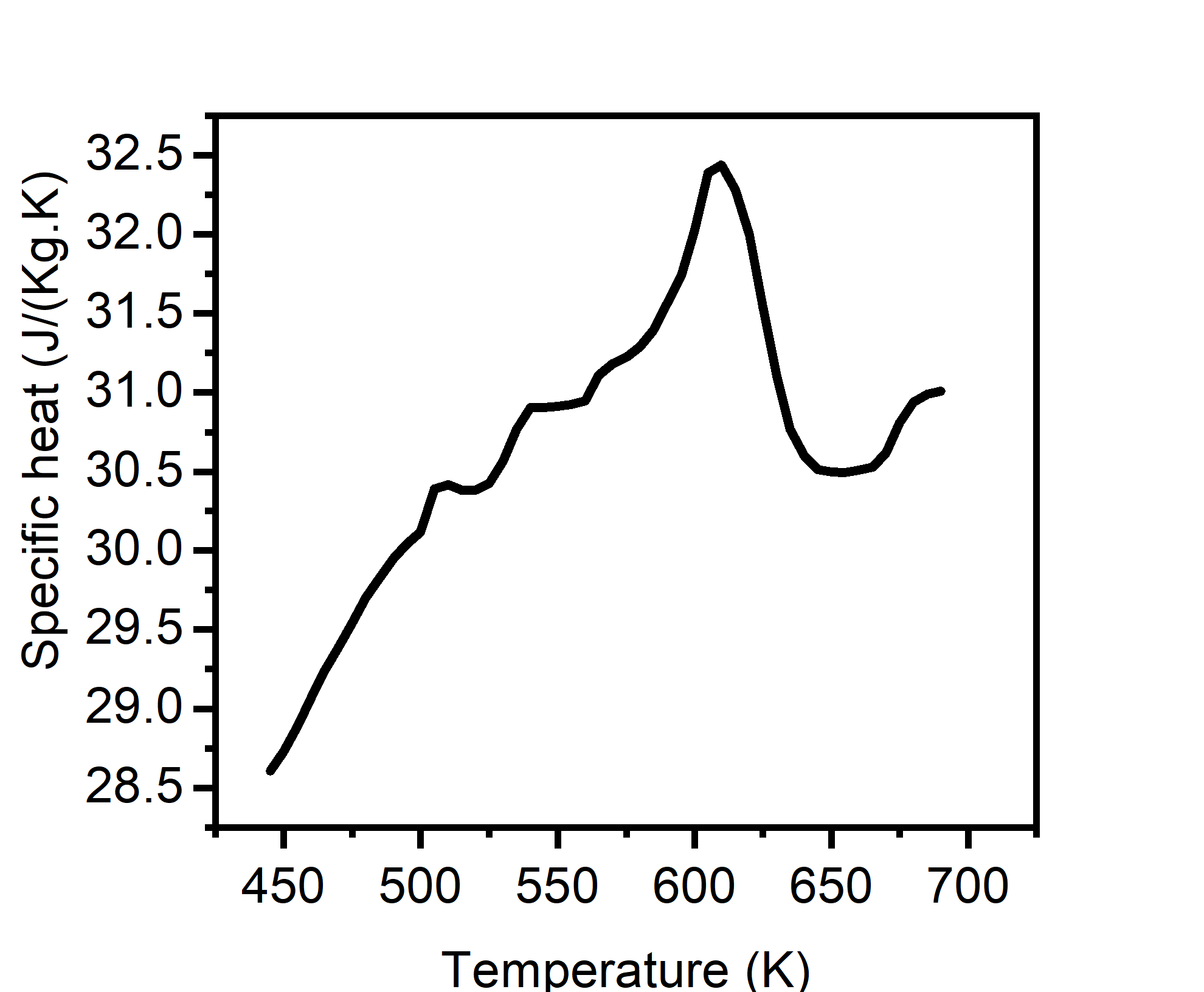}
\caption{Variation of specific heat of GST with temperature from \cite{zalden2014specific}}.
\label{FigS4}
\end{figure}

\begin{figure}[htbp]
\centering
\includegraphics[width=1\linewidth,height=7cm]{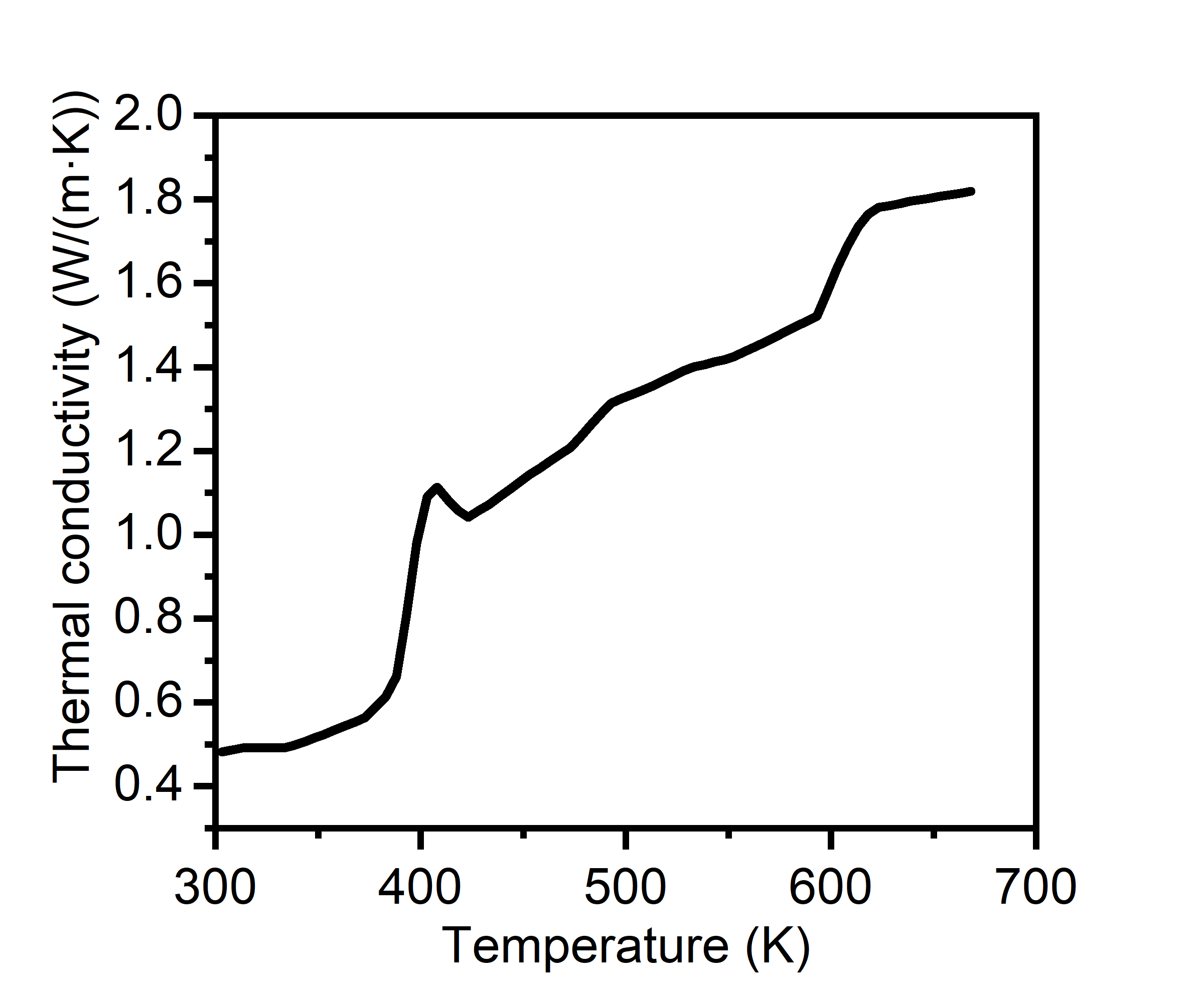}
\caption{Thermal conductivity of GST as a function of temperature derived from \cite{xiong2011low}}.
\label{FigS5}
\end{figure}



\bibliography{bibliography}

\bibliographyfullrefs{bibliography}


\ifthenelse{\equal{\journalref}{aop}}{%
\section*{Author Biographies}
\begingroup
\setlength\intextsep{0pt}
\begin{minipage}[t][6.3cm][t]{1.0\textwidth} 
  \begin{wrapfigure}{L}{0.25\textwidth}
    \includegraphics[width=0.25\textwidth]{john_smith.eps}
  \end{wrapfigure}
  \noindent
  {\bfseries John Smith} received his BSc (Mathematics) in 2000 from The University of Maryland. His research interests include lasers and optics.
\end{minipage}
\begin{minipage}{1.0\textwidth}
  \begin{wrapfigure}{L}{0.25\textwidth}
    \includegraphics[width=0.25\textwidth]{alice_smith.eps}
  \end{wrapfigure}
  \noindent
  {\bfseries Alice Smith} also received her BSc (Mathematics) in 2000 from The University of Maryland. Her research interests also include lasers and optics.
\end{minipage}
\endgroup
}{}

\end{document}